\documentclass[a4paper,11pt]{article}
\pdfoutput=1
\usepackage{latexsym}
\usepackage{feynmp}	%Por algun motivo estos dos codigos hacen funcionar bien los diagramas de feynman y no feynmf
\usepackage{slashed}
\usepackage{cite}

\usepackage{mathtools}
\usepackage[font=small,skip=-20pt]{caption}
\DeclareFontFamily{OT1}{pzc}{}
\DeclareFontShape{OT1}{pzc}{m}{it}{<-> s * [1.10] pzcmi7t}{}
\DeclareMathAlphabet{\mathpzc}{OT1}{pzc}{m}{it}
\usepackage{color}
\usepackage{amssymb}

\begin{document}
 
\title{Double Higgs mechanisms, supermassive stable particles and the vacuum energy}

\author{
Osvaldo P. Santill\'an
\thanks{Electronic addresses: firenzecita@hotmail.com and osantil@dm.uba.ar.}\\
\textit{\small{Departamento de Matem\'atica, Facultad de Ciencias Exactas y Naturales, Universidad de Buenos Aires}}\\
\textit{\small{and CONICET, Ciudad Universitaria, 1428 Buenos Aires, Argentina.}}\\\\
Luciano Gabbanelli
\thanks{Electronic address: lucianogabbanelli@gmail.com.}\\
\textit{\small{Departamento de F\'isica, Facultad de Ciencias Exactas y Naturales, Universidad de Buenos Aires,}}\\
\textit{\small{Ciudad Universitaria, 1428 Buenos Aires, Argentina.}}
}
\date{}
\maketitle

\begin{abstract}
In the present work it is shown that some specific double Higgs like mechanisms may have interesting cosmological applications. A hidden scenario which cast long lived super heavy particles together with an extremely light particle $a$ with mass $m_a\sim 10^{-32}-10^{-33}$eV is presented.  The potential energy of this particle models the vacuum energy density of the universe $\rho_c\simeq 10^{-47}\;\hbox{GeV}^4$. The construction of such scenario is non trivial since the presence of light particles may spoil the stability of the heavy particles. However, double Higgs mechanisms  may be helpful for overcoming this problem. The hidden sector we propose include fermions with masses near the neutrino mass $m_\nu\sim 10^{-2}$eV which arise in terms of a  see saw mechanism. Besides, the super heavy particles acquire a mass due to a double Higgs like mechanism of the order of the GUT scale. The gauge group of the model is $\hbox{SU(2)}_L$ and the scalars of the double Higgs mechanism are not charged under these interactions. The light particle $a$ is the Goldstone boson associated to a Peccei-Quinn like symmetry in the double Higgs model. In addition, the double Higgs mechanism posses another CP odd scalar $A^0$, which acquire a mass of the order of the GUT scale. We show that if there is no direct coupling between $A^0$ and $a$, even in presence of indirect couplings, the $A^0$ particle is long lived an may appear in events above the GKZ bound in present times.

\end{abstract}

\section{Introduction}
In the last years several experimental results have been obtained whose interpretation inarguably demands new physics. One of these observed features is the cosmic acceleration. Since gravity is an attractive force, the velocity of the distant galaxies may be expected to slow down. Contrary to this, the astronomical observations support the fact that this velocity is indeed increasing. Another crucial phenomenon is the discrepancy between the luminous matter of several objects in the universe and their gravitational effects \cite{Rubin,Rubin2}. In fact, there is experimental evidence supporting a flat universe, which implies that the energy density of it should be of the order of the critical one, $\rho_c\simeq 10^{-47}\;\hbox{GeV}^4$ \cite{Carroll}. This scenario does not agree with the contributions corresponding to non relativistic mass density dynamically measured, which are approximately $(0.1-0.3)\rho_c$. Several scenarios have been proposed to explain these results. Some of them postulate the existence of dark matter; i.e an unknown matter sector whose contribution to the energy density compensates the difference between the critical and the observed densities. This hidden sector of particles interacts with the known particles weakly enough not to be detected by current accelerator technology \cite{Salam,Lee,mirror,hidden33,hidden2,wicax,Higgs1,Higgs2,Higgs3,Higgs5,Higgs6}. Furthermore, the acceleration of the universe's expansion suggests the presence of a cosmological constant. If this were to be interpreted as vacuum energy density, then its value would be a considerable fraction of the critical density $\rho_c$. 

The presence of a non zero cosmological constant is an important observation by itself. However, one fact that is striking is that the critical density is not natural from the point of view of the Standard Model. For example, the contribution of the quark condensate $\langle\overline{q}q\rangle$ to the vacuum energy is around 43 orders of magnitude larger than the critical density. For the gluon condensate $\langle G_{\mu\nu}G^{\mu\nu} \rangle$, the discrepancy is even bigger, 44 orders of magnitude. This implies there should be a mechanism for tuning down these contributions to zero. One possibility is the existence of supersymmetry which forces the vacuum energy to be identically zero. However, the presence of a small cosmological constant may signal that supersymmetry is broken; the spontaneously broken supersymmetric theories give contributions of at least 55 orders of magnitude larger than the critical density. For this reason, other scenarios are required to explain this problem. One of the approaches is to postulate that the vacuum energy does not gravitate by some unknown reason and that the effect of a cosmological constant is imitated by the so-called dark energy. The latter is an exotic fluid with negative pressure which drives the cosmic acceleration. Examples of this are the quintessence scenarios \cite{quintesence}. In these models the vacuum energy is associated with a slowly rolling scalar field $\varphi$ under the influence of a nearly flat potential $V(\varphi)$. The nearly flat condition ensures that $\varphi$ is not at the minimum of its potential $V(\varphi)$ at present times. As a consequence, the vacuum energy is a temporary effect which disappears for large times.

The quintessence scenarios are able to describe the current energy density, but the assumption that the vacuum energy does not participate in the gravitational interaction is still to be studied further. If this assumption is relaxed, then other type of scenarios should be introduced, such as the so-called cancellation (or adjustment) mechanisms \cite{dolgov,dolgov2,urban,emelyanov,emelyanov1,emelyanov2}. These models assume the presence of an initial gravitating energy density, together with an unknown component which contributes to this density with opposite sign, in such a way that for large times the total energy density becomes very small. The cancellation mechanisms based on scalar fields suffer some potential problems, as noticed already in \cite{dolgovv,weinberg}. Nevertheless, there are some scalar models that may avoid these complications although they usually take place in modified theories of gravity \cite{dolgov2,urban}. The cancellation mechanisms for higher spin fields do not have these problems, since by construction they predict a very small vacuum energy. The possibility of them modeling the critical energy density without screening the Newton gravity constant is not excluded \cite{emelyanov,emelyanov1,emelyanov2}.

The present work assumes the existence of a suitable adjustment mechanism, capable of lowering down the energy densities contributions mentioned above. Bearing this assumption in mind, the present energy density will be modelled in terms of a very light axion $a$, with a mass of $m_a\sim 10^{-32}-10^{-33}\;\hbox{eV}$. This axion arises due to a symmetry breaking mechanism in a hidden sector corresponding to an $\hbox{SU(2)}_\mathrm{L}$ gauge interaction whose unification with QCD is produced at a very large energy scale of the order of the GUT or even of the Planck energy. This theory is confining at a very low scale, of the order of the light neutrino mass $m_\nu\sim 10^{-2}-10^{-3}\;\hbox{eV}$. The model posses an approximate Peccei-Quinn like particle, which is violated by the small masses of some hidden fermions. As a consequence the axion acquires the mass $m_a\sim 10^{-32}-10^{-33}\;\hbox{eV}$. The large scale of the spontaneously symmetry breaking implies that a hidden sector contains superheavy particles. These particles could be formed in an early stage of the universe and their large mass values may suggest that their mean lifetime is too short to be present in the actual universe. However, it will be argued below that this conclusion is not necessarily true. We will show that an scenario with a supermassive sector  can be cast in the model, with mass of at least $10^{15}\;\hbox{GeV}$, in such a way that there appears an stable CP scalar  $A^0$, with mean lifetime equal or greater than the age of the universe, $\tau\sim 10^{11}\;\hbox{yrs}$. As a result, this sector can arrange events above the Greisen-Zatsepin-Kuzmin (GKZ) limit \cite{GKZ}, which is a very interesting fact to be taken into account.

The stability issue described above is non trivial. In fact, the task of accommodating such stable particle in presence of a very light Goldstone particle is a difficult one, since a direct or indirect coupling between the heavy particle can induce a fast decay channel which spoils its stability. Thus, the problem is a delicate one. In the present work these difficulties are described in certain detail, however it is argued below that in some restricted cases stability can be warranted.

\vspace{10pt}\noindent\textbf{Outline:} The organization of this paper is as follows. In section 2 general properties of QCD axion models are discussed which are required for the construction of our model. In section 3 the Lagrangian for the hidden sector is presented. In section 4 we describe in detail the couplings of the particle $A^0$  with the other particles of the model. Section 5 contains an estimation of the mean lifetime of one of the massive CP odd scalar particle composing this sector, and it is shown that this particle is stable. Section 6 contains a discussion of the results.

\section{A brief review of axion mechanisms in QCD}

The present section gives a description of some axion models of QCD with a two-folded motivation. On the one hand, the axion scenarios will be useful to construct our model. On the other, they can be considered as an example of a cancellation mechanism in which an extremely small parameter $\theta_{eff}$ is interpreted in terms of a dynamical component, the axion. This cancels the effect of the bare parameter $\theta$, resulting in an extremely small effective value $\theta_{eff}$.

As is well known in ordinary QCD, the $\theta$ term associated with the instantons solutions and related quantum effects of the theory \cite{belavin,thooft}
\begin{equation}
	\mathcal L_{\theta}= \frac{\theta}{32\pi^2} G^a_{\;\mu\nu}\widetilde{G}^{a\;\mu\nu}  \;,
\end{equation}
violate CP invariance when the fermions of the theory are massive. In the latter expression, $G^a_{\;\mu\nu}$ is the gluon strength field and $\widetilde{G}^{a\;\mu\nu}$ its dual expression. For massless QCD instead, the chiral transformation
\begin{equation}
	\label{rut}
	\psi\to e^{i\gamma_5 \alpha}\psi
\end{equation}
on the fermions wave functions $\psi$ of the theory, is a classical symmetry of the Lagrangian. On the contrary, at a quantum level there is an anomaly in the chiral current $J^{\mu5}$ given by
\begin{equation}
	\label{anomu}
	\partial_{\mu}J^{\mu5}= \frac{g^2}{16\pi^2} G^a_{\;\mu\nu}\widetilde{G}^{a\;\mu\nu}  \;.
\end{equation}
For this reason, if the fermions were massless, the chiral transformation would modify the $\theta$ parameter in the following way:
$$
\theta\to \theta-2\alpha  \;.
$$
This means that for massless QCD all the theories with different $\theta$ would be equivalent. Thus, it is the mass term of the fermions which spoils the chiral symmetry and simultaneously the CP invariance.

The value of $\theta$ is not fixed by the theory itself and should be determined by the experiments. The experimental known bound is $\theta<10^{-9}$. This value does not satisfy the majority of the scientific community, which regards the introduction of such small parameter in the theory as unnatural. For this reason in \cite{axion1} an alternative to explain this lack of naturalness problem was introduced. They consider the $\theta$ parameter as a dynamical field, the axion, which runs to the value zero regardless of its initial value. The effective Lagrangian describing the axion $a$ and its interaction with the gluons is
$$
\mathcal L_{eff}=\mathcal L_{QCD}+\mathcal L_k(a)+\left(\theta+\frac{a}{f_a}\right)\frac{\alpha_s}{8\pi}G^a_{\mu\nu} \widetilde{G}^{a\mu\nu}  \;,
$$
where $f_a$ stands for the axion constant; $\mathcal L_{k}(a)$ its kinetic term and $\alpha_s$ a constant. Grouping the $\theta$ term with the coupling of the axion to the boson gauge fields of QCD, it is possible to redefine an effective $\overline{\theta}$ term
\begin{equation*}
	\overline{\theta}=\theta+\frac{a}{f_a}  \;.
\end{equation*}
After shifting the field as follows $a\to a-f_a \theta$, the $\theta$ parameter can be discarded. This implies the theory will be CP invariant if we can find some mechanism which forces the axion to take the value $a=0$. This is in fact precisely what happens, since the axion is under the influence of an effective quantum potential $V(a)$, due to the effect of the quarks and gluons inside the Feynman path integral whose minimum is $a=0$. This potential is given by
\begin{equation*}
	\exp\left[-\int d^4x V(a)\right]= \int DA_\mu\prod_{i}Dq_iD\overline{q}_i \exp\left[-\int d^4x\left(\mathcal L_{QCD} +\frac{\alpha_s}{8\pi f_a}a\;G^a_{\mu\nu}\widetilde{G}^{a\mu\nu}\right)\right]  \;,
\end{equation*}
and its explicit expression has been presented in \cite{Kim1987} as follows
\begin{equation}
	\label{poto}
	V(a)\sim f^2_{\pi} m^2_{\pi} \bigg[1-\cos\left(\frac{a}{f_a}\right)\bigg]  \;;
\end{equation}
where $f_{\pi}$ and $m_{\pi}$ are the pion's coupling constant and mass respectively. Expression (\ref{poto}) implies that the minima is at $a=0$, solving the CP problem. At the same time, the lack of flatness of the potential causes the axion to develop a mass of order
\begin{equation}
	\label{rev}
	m_a\sim \frac{f_{\pi}m_{\pi}}{f_{a}}  \;.
\end{equation}
Although results given above take into account the color interaction, they should be supplemented with the CP violating terms of the weak interaction. This produces a very tiny but non zero value $\theta_{eff}$.

There are several axion scenarios discussed in the literature \cite{axion11,axion12,axion3,axion4,axion6,axion7,axion7bis,axion5,SVZ1980,kim1979}. In some of them, the axion does not interact directly with the ordinary quarks, but through the gluons and the coupling $a G \widetilde{G}$ is interpreted as an effective interaction. In the diagram presented in figure \ref{axion_gluon}, the triangle is composed by a hidden heavy quark $Q$, giving rise to an effective interaction of the form $f^{-1}_{a}\;a G \widetilde{G}$. In this case, the axion parameter $f_a$ is a function of the mass $m_Q$ of this new quark.

\begin{figure}[!t] % [!htb] la deja en el medio
	\begin{center}
		\begin{tabular}{cccccccccccccccc}    %Adds a bunch of centered Columns
			\begin{fmffile}{axion_gluon} 	%one.mp will be created for this feynman diagram
				\fmfframe(6,42)(6,42){ 	%Sets dimension of Diagram
					\begin{fmfgraph*}(220,124) %Sets size of Diagram
						\fmfleft{i}	%Sets there to be 2 sources
						\fmfright{o1,o2}    %Sets there to be 2 outputs
						\fmf{scalar,label=$a$,label.side=left}{i,iv3}
						\fmf{fermion,label=$Q$,label.side=right}{iv1,iv2,iv3,iv1}
						\fmf{gluon,label=$G$,lab.dist=0.05w}{iv1,o1}
						\fmf{gluon,label=$G$,label.side=left}{iv2,o2}
						\fmffixed{(0,.7h)}{iv1,iv2} 
					\end{fmfgraph*}
				}
			\end{fmffile}
		\end{tabular}
		\caption{\textsf{Diagram of the axion effective coupling to the gluons of QCD through the new heavy quark $Q$. As in the Standard Model, the gluons interact as usual with the ordinary quarks.}}
		\label{axion_gluon}
	\end{center}
\end{figure}

In the upcoming analysis, we will consider as presented in \cite{kim1979,SVZ1980}, the addition of the following terms corresponding to the wave function $\psi$ of a supermassive quark $Q$ coupled to a scalar field $\varphi$, to the Lagrangian of the theory of QCD
\begin{equation}
\label{ax}
\mathcal L_{add}= i\overline{\psi}\slashed{D}\psi -(\delta \overline{\psi}_R \varphi \psi_L +\delta^\ast \overline{\psi}_L \varphi^\ast \psi_R) +(\partial_{\mu}\varphi^\ast)(\partial_{\mu}\varphi) +m^2\varphi^\ast\varphi -\lambda(\varphi^\ast\varphi)^2  \;.
\end{equation}
The first term $i\overline{\psi}\slashed{D}\psi$ includes the kinetic energy of the new quark and its coupling with the gluons; the parameters $\lambda$, $m$ and $\delta$ are to be determined. Since the new scalar field shows a non-vanishing vacuum expectation value $\mid<\varphi>\mid=m/\sqrt{2\lambda} \equiv\varphi_0$, it acquires a mass $\sqrt{2}m$. Furthermore, the last-mentioned v.e.v. rises a mass for the heavy quark given by $m_\psi=\delta\varphi_0$. In addition, there is a massless pseudoscalar $a$ defined by
\begin{equation}
\label{aha}
\varphi=(\varphi_0+\rho) \exp\left(i\frac{a}{\varphi_0\sqrt{2}}\right) \;.
\end{equation}
The $\rho$ field describes the radial excitations and $a$ the angular ones. The pseudoscalar $a$ is identified with the axion and it is the Goldstone boson associated to the breaking of the $\hbox{U(1)}_{PQ}$ symmetry transformation of the Lagrangian (\ref{ax}):
\begin{equation}
\label{mic}
\psi\to e^{i\gamma_5 \alpha}\psi  \; ,\qquad\qquad \varphi\to e^{-2i\alpha}\varphi  \;,
\end{equation}
This axion field does not interact at the tree level with the light quarks and gluons, but acquires an effective interaction with the last field due to the diagram shown in figure \ref{axion_gluon}. The resulting interaction, then, is determined by the quark loop and thus, the effective Lagrangian has the form
\begin{equation*}
	\frac{\alpha_s}{8\pi\sqrt{2}\varphi_0}a G^a_{\mu\nu}\widetilde{G}^{a\mu\nu}  \;.
\end{equation*}
From here it follows that the axion coupling constant is related to the vacuum expectation value according to $f_a=\sqrt{2}\varphi_0$. This result implies that the mass of the quark $\psi$ is proportional to $f_a$; so the heavier the quark is, the lighter the axion will be.

Whether an axion mechanism which solves the CP problem in QCD exists or not is an open question, but the particular scenario presented above will be helpful for constructing our model.

\section{The vacuum energy density as an axion like particle}
In order to describe the present energy density of the universe, it is of interest to find mechanisms giving rise to extremely light particles whose characteristic time $t_{m}$ is of the order of the universe's age. As is well known in classical cosmology, the Hubble parameter is related to the critical density $\rho_c$ and to the Newton constant $G_N$ by means of the Friedmann classical equation
\begin{equation}
	\label{hub}
	H^2=\frac{8\pi}{3}G_N \rho_c  \;.
\end{equation}
Furthermore, the Hubble constant today can be parametrized as
\begin{equation}
	\label{hubact}
	H_0\simeq  \frac{M^2}{M_{Pl}} \;
\end{equation}
with $M\sim 10^{-2}-10^{-3}\;\hbox{eV}$, which is a value not far from the mass of a light neutrino $m_\nu$. This equation is expressed in natural units $\hbar=c=1$; making explicit the factor $c/\hbar^2$ on the right hand side, the equation manifests its quantum behavior\footnote{In fact, interesting relations between $H_0$ and the pion mass $m_\pi$ have been pointed out in \cite{Calogero} and worked out further in \cite{Estrada}.}. The left hand side of (\ref{hubact}) is the result of experimental cosmological observations, while the right one involves the lightest and the heaviest scale known in physics. A dynamical interpretation of this numerical relation may be given by modeling the energy density of the universe with an extremely light particle $a$ with mass $m_{a}\simeq H_0$, which is assumed to represent a large fraction of the dark energy. The numerical value of (\ref{hubact}) in natural units is
\begin{equation}
	\label{hub3}
	m_{a}\sim 10^{-32}-10^{-34} \;\hbox{eV} \;.
\end{equation}

\noindent The characteristic time scale, $t_{a}=1/m_{a}\sim 10^9-10^{10}\;\hbox{yrs}$ for such particle is close to the estimated age of the universe. Thus, if such particle exists, together with a suitable component which cancels the QFT vacuum contribution, it will be the predominant component of the energy of the universe.

A dynamical interpretation of these relations can be found in terms of a hidden sector of the type described in section 2. Scenarios like those, implies the existence of a pseudoscalar $a$, the axion, under the influence of an effective potential of the form
\begin{equation}
\label{fpot}
V(a)\sim M^4\bigg[1-\cos\bigg(\frac{a}{f_{a}}\bigg)\bigg]  \;,
\end{equation}
which is analogous to (\ref{poto}). In the following, the notation $a$ for this hidden axion will be employed, although this pseudo scalar should not be identified with the  QCD  invisible axion. The quantity $M^4$ has energy density units and can hence be considered as $M\sim (\rho_c)^{1/4}$. The axion mass $m_{a}$ is related to the scale $M$ appearing in (\ref{fpot}) through the relation
\begin{equation}
\label{isi}
m^2_{a}\sim\frac{M^4}{f^2_{a}}\sim 10^{-64}\; \hbox{eV}^2  \;.
\end{equation}
In the last step (\ref{hub3}) was taken into account. The combination of $M\sim (\rho_c)^{1/4}$ and (\ref{isi}) gives a value for $f_{a}$ of the order of the Planck mass, that is, $f_{a}\sim 10^{19}\;\hbox{GeV}$. In fact, $(\rho_c)^{1/4}$ is not far from a light neutrino mass $m_\nu$. This values has already been considered in the context of the solar neutrino problem \cite{frieman,frieman3} and it is also taken into account in \cite{caldwell}. In any case, a field with such characteristics is frozen by the Hubble constant along almost all the cosmic history and at the present times will be rolling to the minimum of its potential \cite{quv}. The initial value of the field, of course, determines the dynamics today. Although this may imply a fine tuning for the initial conditions, the fact that the potential is periodic softens the problem, since there is an appreciable fraction of the range $(0,2\pi a)$ of $a$ which gives a viable dynamic at the present era.

The next goal is to construct a scenario with these characteristics. As a preliminary step, consider a hidden sector with two types hidden fermions, which interacts under a hidden  $\hbox{SU(2)}$ gauge interaction. 
The first type of fermion has two components
$$
f_i=\left(
\begin{array}{cc}
  f_{i1}      \\
  f_{i2}      
\end{array}
\right)  \;.
$$
Here $i=1,..,n$ is the number of such particles. The gauge symmetry acts as
$$
f_i\to \exp(i\sigma^a \alpha_a(x))f_i.
$$
The hidden symmetry invariance imply the presence of three mediating bosons $G_i$ with  $i=1,2,3$. The mass $m_{f_i}$ will be assumed to be tiny, of the order of the neutrino mass $m_\nu$. The second fermion is
$$
F=\left(
\begin{array}{cc}
  F_1     \\
  F_2    
\end{array}
\right)  \;,
$$
and its mass $m_F$ will be assumed to be of the order of the $M_{GUT}$ or even $M_{Pl}$. These particles constitute a hidden sector, and are not identified with states of the Standard Model.
The mechanism to be described above bears an analogy with the KSVZ axion \cite{kim1979,SVZ1980}. The fermion F plays the role of the heavy quark Q in the KSVZ scenario and the light fermions $f_i$ the role of the ordinary quarks of that model. The fermions masses are obtained through a spontaneously symmetry breaking mechanism with two different Higgs like particles $\Phi_1$ or $\Phi_2$. For instance, the part of the lagrangian corresponding to $F$ and $\Phi_2$ is
\begin{equation}
\label{axl}
\mathcal L_{add}= i\overline{F}\slashed{D}F -(\delta \overline{F}_R \Phi_2F_L +\delta^\ast \overline{F}_L \Phi_2^\ast F_R) +(\partial_{\mu}\Phi_2^\ast)(\partial_{\mu}\Phi_2) +m^2\Phi_2^\ast\Phi_2-\lambda(\Phi_2^\ast\Phi_2)^2  \;.
\end{equation}
The analogous lagrangian follows for $f$ and $\Phi_1$. There are two pseudo scalars $a$ and $b$ are defined through
\begin{equation*}
	\Phi_1=(v_1+h_1) e^{i \frac{b}{f_b}} 
	\qquad \hbox{and} \qquad 
	\Phi_2=(v_2+h_2) e^{i \frac{a}{f_a}}  \;.
\end{equation*}
Our assumption that $\Phi_2$ does not give mass to any gauge mediating boson implies that $a$ can not be gauged away, so it become physically relevant. The field $b$ is irrelevant for the following discussion. It may be gauged away by giving mass to other interactions or not, but this will not be essential. The gauge lagrangian is
$$
L_g=\frac{1}{4\pi}G_{\mu\nu}\cdot G^{\mu\nu}+\frac{\theta_h}{4\pi}G_{\mu\nu}\cdot\widetilde{G}^{\mu\nu}  \;.
$$
The mass terms for the fermions  are all of Dirac type and $\theta_h$ is due to the effect of the instantons of the non-Abelian theory. 

The $SU(2)$ interaction at this point is generic, and is not to be confused with any interaction of the Standard Model. But it will be further specified by the request to give rise to a light axion such as (\ref{isi}). Examples of interaction of this type can be found in the literature about schizons  \cite{frieman,frieman3}. Following those references consider that the $\hbox{SU(2)}$ gauge interaction corresponds to $n_f$ fermion flavors, and unifies with ordinary QCD at a very large scale $\widetilde{M}$ of the order of $M_{GUT}$ or $M_{Pl}$. Then, renormalization group arguments show that this interaction confines at a scale $\Lambda_h$ given by
\begin{equation}\label{upnight}
	\Lambda_{h}\sim \widetilde{M} \left(\frac{\Lambda_{QCD}}{\widetilde{M}}\right)^{\bigg(\frac{33-2n_f}{22-n'_f}\bigg)} .
\end{equation}
When $n'_f=4$ and using that $n_f=6$ then, taking  into account that $\Lambda_{QCD}\sim 0.1\;\hbox{GeV}$, it follows that when $\widetilde{M}$ is of the order of $M_{GUT}$ or even of the Planck scale, then  the confining scale is comparable to $\Lambda_h\sim 0.01\;\hbox{eV}$. Clearly this scale resembles the mass of a light neutrino $m_\nu$. The $\theta_h$ term is cancelled by the Goldstone pseudo scalar $a$ as in ordinary QCD axion models. Moreover, in these models, the mass of the axion is usually given by
\begin{equation*}
	m^2_a\sim \frac{m_{f}\; \langle
		\overline{f\raisebox{7pt}{}}f \rangle} {f^2_a}  \;,
\end{equation*}
where it is a reasonable to assume that the mean values $\langle\overline{f\raisebox{7pt}{}}f\rangle$ are close $\Lambda^{3}_h$. Note that both this scale and $m_f$ are of the order of a light neutrino. The axion thus gets a small mass, of the order of
\begin{equation}\label{vives}
	m_a\sim\frac{m_\nu^2}{f_a}\sim 10^{-32}-10^{-33} \;\hbox{eV}  \;.
\end{equation}
Henceforth, the condition (\ref{isi}) is reproduced by this scenario. This means that the Compton wavelength associated to this mass is comparable to the radius of the observable universe and that this axion $a$ describes a considerable fraction of the energy density of the present universe.

There exist recent models of this type, examples are \cite{quv} and \cite{nilles}-\cite{nilles2}.
 
\section{The super-heavy sector of the proposed model}
\subsection{The lagrangian and the role of the double Higgs mechanism}
The next step is to modify the model discussed above in order to include stable superheavy particles. This is a non trivial task, since the presence of light particles, with masses of the order or much below the neutrino mass may spoil the desired stability. However, double Higgs mechanisms may be helpful for circumventing the problem, as it will be explained below. Such stable heavy particles may act as ultra massive dark matter at present times. Our goal is to present the lagrangian which incorporates these characteristics first and to explain how it works afterwards.  

First of all, it is assumed the existence of an $\hbox{SU(2)}_L$ gauge interaction.The corresponding mediating bosons will be denoted as $G^i_\mu$ with $i=1,2,3$, and will acquire a mass due to a spontaneous symmetry breaking mechanism.  The particles of the ordinary Standard Model are assumed to be charged under the $\hbox{SU(2)}_L$ interaction as well. This is requested in order to describe high energy decays of a hidden super heavy particle into ordinary ones, otherwise the hidden sector will be sterile. With these ideas in mind, the lagrangian we propose is
\begin{equation}\label{mod2}
\mathcal L=\mathcal L_{schizon}+i\sum_{i=j}^2\overline{F}_{jL}\slashed{D} F_{jL} + i\sum_{j=1}^2\overline{F}_{jR}\slashed{\partial} F_{jR}+
|\nabla_\mu \Phi_1|^2+|\nabla_\mu \Phi_2|^2 +\frac{1}{4\pi}G_{\mu\nu}\cdot G^{\mu\nu} \end{equation}
$$+\frac{\theta_c}{4\pi}G_{\mu\nu}\cdot\widetilde{G}^{\mu\nu}
+V(\Phi_1, \Phi_2)+\mathcal L_{sm}(X_i, G_\mu),
$$
where $X_i$ are some particles of the ordinary Standard Model. The lagrangian $\mathcal L_{schizon}$ will contain an axion which will model a considerable fraction of the vacuum energy, and will be described in detail in the following sections. The following discussion is focused on the remaining part of the lagrangian.

The covariant derivative $\nabla_\mu$ includes  the $G_\mu^i$ fields. The mass terms for the $F_1$ and $F_2$ are obtained in terms of a double Higgs mechanism. The potential $V(\Phi_1, \Phi_2)$ generating their masses  corresponds to a doublet potential, described in detail in \cite{georgi,hunter}. Its expression is given by
$$
V(\Phi_1, \Phi_2)= Y_1 \Phi_1^\dagger \Phi_1+ Y_2 \Phi_2^\dagger \Phi_2
+[Y_3 \Phi_1^\dagger \Phi_2+h.c]
$$
\begin{equation}\label{poton}
+ Z_1(\Phi_1^\dagger \Phi_1)^2 + Z_2(\Phi_2^\dagger \Phi_2)^2
+Z_3(\Phi_1^\dagger \Phi_1)(\Phi_2^\dagger \Phi_2)
+Z_4( \Phi_1^\dagger \Phi_2)(\Phi_2^\dagger \Phi_1)
\end{equation}
 $$
 + \bigg[Z_5 (\Phi_1^\dagger \Phi_1)^2
+\bigg(Z_6 (\Phi_1^\dagger \Phi_1)
+Z_7 (\Phi_2^\dagger \Phi_2)\bigg)
\Phi_1^\dagger \Phi_2+h.c \bigg]\,,
$$
with $Z_i$ and $Y_i$ the complex parameters of the potential. One of the reasons for the choice of a double Higgs mechanism is the presence of a massive CP odd scalar particle $A^0$ which is always present in these models \cite{hunter}.  The stability issues to be discussed further on are more tractable for CP odd particles \cite{kim1979}, so this election is for simplicity. The particle $A^0$ is given by \cite{hunter}
\begin{equation}\label{acero}
A^0=\sqrt{2}(-\sin\beta \mathcal Im \Phi_1+\cos\beta \mathcal Im \Phi_2),\qquad \tan \beta=v_2/v_1,
\end{equation}
with $v_i=<\Phi_i>$. Its mass is of the order of the scale of symmetry breaking $m_{A^0}\sim Z f_a$, with $Z$ a combination of the parameters of the model.  If this breaking takes place at a large energy scale, this particle becomes very heavy. This is a candidate for a super heavy stable particle, which is one of the features this work is aimed for.

For all these models, there is a parameter $\xi$, constructed in terms of the parameters $Z_i$ and $Y_i$, such that the model is CP violating unless $\xi=0$. The last situation is assumed in the following description. 

With the elements introduced above, the mass term $L_m(F_i, \Phi_1, \Phi_2)$ written in (\ref{mod2}) can be described as follows. The fermions  $F_1$ and $F_2$ are extended to a doublet $Q$. There exist several possible mass terms in the double Higgs model, an example is the following
\begin{equation}\label{osval}
L_{m}=-\gamma^\ast \overline{Q}\Phi^c_1 F_1 -\gamma \overline{Q}\Phi_1 F_2-\lambda^\ast \overline{Q}\Phi^c_2 F_1 -\lambda \overline{Q}\Phi_2 F_2+h.c,
\end{equation}
with $\Phi_i^c=i\sigma_2 \Phi^\ast$. It should be emphasized however, that the doublet $Q$ is \emph{not suggesting the presence of an additional hidden flavor interaction}. 
Another possibility is giving mass terms of the form
\begin{equation}\label{osvaldito}
L_{1m}=-\gamma^\ast \overline{Q}\Phi^c_1 F^{(1)}_{1} -\gamma \overline{Q}\Phi_1 F^{(2)}_1-\lambda^\ast \overline{Q}\Phi^c_2 F^{(1)}_1 -\lambda \overline{Q}\Phi_2 F^{(2)}_1+h.c,
\end{equation}
and the analogous for the fermion $F_2$. The super index $(i)$ is indicating the i-th "color" component of the fermion $F_1$ or $F_2$. Clearly, with the mass term (\ref{osval}) all these components
have the same mass and the gauge interaction imitates partially a color. For (\ref{osvaldito}) the mass of the components $F_{1i}$ with $i=1,2$ are different and the interaction looks more like a flavor.
For the stability matter discussed below, both terms are allowed.

It is important to mention that in these scenarios there exist a Goldstone boson $G^0$, which is related to certain $U(1)$ global symmetry which involves the fermions $F_i$ and $\Phi_i$ \cite{hunter}. This symmetry is a generalization of the Peccei-Quinn one for the double Higgs mechanism. However, this mechanism is also giving mass to the mediating bosons $G_\mu$ and therefore, the boson $G^0$ can be gauged away. This can be seen by standard arguments. The double Higgs model has also some scalar states, but the choice $\xi=0$ forbids any coupling between these scalars and the particle $A^0$. This symmetry forbids indirect couplings between these states as well \cite{hunter}. Extensive formulas about the double Higgs mechanism in the Standard Model can be found in \cite{hunter3}, and we refer the reader there for further information.

 \subsection{The stability matters for the  $A^0$ particle}

 The next aspect to be discussed is the stability of the odd particle $A^0$. This is clarified by studying its coupling with the other particles of the model. These couplings are described in the references \cite{hunter}-\cite{gunion}. In particular, the formula (4.21) of the reference \cite{hunter} shows that a typical coupling between $A^0$ and the fermions of the theory is given by
$$
\mathcal L_{A^0 F_i \overline{F}_i}\sim c A^0 \overline{F}_i\gamma_5 F_i,
$$
with $c= \lambda$ or $c=\gamma$, the coupling constants defined in (\ref{osval}). The requirements for stability is then that the mass $m_{A^0}$ is the smaller than the $2m_{F}$ and smaller than  $2m_{G^i}$. However, it will be assumed that this particle has a large mass value, which implies the same property for the $G^i$ and the $F_i$ particles. 

The inequality $m_{A^0}<2m_{F_i}$ just mentioned is required to kinematically forbids  the decay $$A^0\to \overline{F}_i +F_i.$$If these decays were allowed, then $A^0$ would be short lived. The condition  $m_{A^0}<2m_{G^i}$ forbids the decay of $A^0$ into two $G^i$ bosons through the ABJ anomaly diagram, analogous to the one presented in Figure \ref{axion_gluon}, in which the internal triangle is composed by any of the $F_i$ fermions. This diagram also gives a short mean lifetime, but the condition $m_{A^0}<2m_{G^i}$ avoid this problem.

The other states of the double Higgs model are scalars \cite{hunter}. Clearly, a direct or indirect coupling between $A^0$ and these scalar states is forbidden by the CP invariance.

\begin{figure}[!b]
	\begin{center}
		\begin{tabular}{cccccccccccccccc}    %Adds a bunch of centered Columns
			\begin{fmffile}{axion_G_q} 	%one.mf will be created for this feynman diagram
				\fmfframe(6,42)(6,42){ 	%Sets dimension of Diagram
					\begin{fmfgraph*}(220,124) %Sets size of Diagram
						\fmfleft{i}	%Sets there to be 2 sources
						\fmfright{o1,o2}    %Sets there to be 2  outputs
						\fmflabel{$q$}{o1}
						\fmflabel{$q$}{o2}
						\fmf{scalar,label=$a$,label.side=left}{i,iv3}
						\fmf{fermion}{ov1,o1}
						\fmf{fermion}{o2,ov2}
						\fmf{fermion,label=$\psi$,label.side=right}{iv1,iv2,iv3,iv1}
						\fmf{fermion,label=$q$,label.side=left}{ov2,ov1}
						\fmf{gluon,label=$G$,lab.dist=0.05w}{iv1,ov1}
						\fmf{gluon,label=$G$,label.side=left}{iv2,ov2}
						\fmffixed{(0,.7h)}{iv1,iv2} \fmffixed{(0,.7h)}{ov1,ov2}
					\end{fmfgraph*}
				}
			\end{fmffile}
		\end{tabular}
		\caption{\textsf{Diagram which describes the effective coupling between the KSVZ invisible axion and the QCD quarks $q$. The triangle lines corresponds to a hidden massive quark $Q$, which is a singlet under the electroweak interaction.}}
		\label{axion_G_q}
	\end{center}
\end{figure}

With the couplings described in the previous section, it follows after drawing the allowed Feymann diagrams for the decay of $A^0$, that the lowest order diagram is of the type of the Figure \ref{higgs_quark}. The internal boson lines can be related to the $\hbox{U(1)}$ or the $\hbox{SU(2)}$ interaction. The products of the decay kinematically allowed are the light hidden fermions with masses $m_\nu$ or fermions of the ordinary Standard Model, with masses of the order of  $m_q\sim$MeV. Let us assume first that the products of the decay are the light hidden fermions, with masses $m_{\nu}$. At first sight, the diagram of Figure \ref{higgs_quark} may give a short lifetime for $A_0$ since smaller the mass of the decaying product is, the more probable the decay becomes. Fortunately, this apparently intuitive argument does not hold here. This extremely important issue can be clarified by considering some known cases in the literature.

\subsection{Comparison with an axion mechanism due to Kim}

 In order to understand the behavior of the diagram in Figure \ref{higgs_quark} with respect to the mass parameters of the model, consider first the situation in which the masses corresponding to the fermion triangle are very large values $m_{F}\to \infty$. Then it is reasonable to expect that, in this limit, the decay probability will decrease to zero and therefore the mean lifetime $\tau$ will becomes infinite. In other words, it is expected that
\begin{equation}
	\label{con1}
	\lim_{m_{F}\to \infty}\tau\to \infty  \;.
\end{equation}
On the other hand, it is also feasible  that when $m_{A^0}$ increases to $m'_{A^0}>m_{A^0}$, the decay will becomes more probable and the mean lifetime will decreases, that is
\begin{equation}
	\label{con2}
	m_{A^0}<m'_{A^0}\qquad \Rightarrow\qquad \tau(m_{A^0})>\tau(m'_{A^0})  \;.
\end{equation}
Instead, the behavior of the amplitude with respect to the mass $m_\nu$ of the products of the decay is more involved. A decrease of this mass is equivalent to an increase of the masses $m_F$ and $m_{A^0}$ with their ratio $R=m_{F}/m_{A^0}>1$ fixed. The increase of the mass of the Higgs results in a shorter lifetime. However (\ref{con1}) implies that the increment on the loop triangle mass enlarges the lifetime of the boson. As a consequence, there is a competence between those effects in the fixed ratio limit, and it is not clear which of the two effects prevails, if any. 

Fortunately , the work of Kim  \cite{kim1979} considers the diagram in Figure \ref{axion_G_q}, which bears an strong analogy to the one in Figure \ref{higgs_quark}. For Kim, the decaying particle is a QCD axion $a$, the triangle is composed by a heavy quark $Q$ and the gauge mediators are ordinary QCD gluons.  Kim's diagram induces an effective coupling between the decaying axion and the ordinary quarks whose schematic behavior is
\begin{equation}
\label{kimon}
g_{eff}\simeq g_{\mathrm{QCD}}^4\frac{m_q}{m_Q}  \ln\left(\frac{m_Q}{m_q}\right)  \;.
\end{equation}
Here $m_Q$ is the mass of the heavy quark and $m_q$ is the mass of any ordinary quark resulting from the decay. The decay rate is then
\begin{equation}\label{ohohoho}
\Gamma_a=\alpha_{\mathrm{QCD}}^4\bigg(\frac{m_q}{m_Q}\bigg)^2 m_a   \ln^2\left(\frac{m_Q}{m_q}\right).
\end{equation}
In this expression, it is found that $g_{eff}\to 0$ when $m_Q\to \infty$ which is what intuition suggests. However it is also be seen that $g_{eff}\to 0$ when the masses $m_q\to 0$, which is opposite to what a first intuition suggest.
These conditions imply that the mean lifetime of the axion becomes infinite in this limit, since there is no coupling and thus no decay. In other words, the lighter the ordinary quarks are, the larger the mean lifetime of the axion becomes. This fact is showing that, for CP odd particles, the effect due to the increment o the mass of the triangle is the one which prevails.

The Kim result discussed above has applications related to the present work. First of all, it open the possibility that the main decaying product are fermions of the ordinary Standard Model, with typical masses $m_f\sim$MeV, instead of the hidden fermions with masses $m_\nu$. The reason is that, as discussed above, $g_{eff}\to 0$ when this mass decreases. Thus, the main diagram for the decay is the one with the heaviest decay products.
This opens the possibility of detecting the decay of the CP odd scalar $A^0$ through high energy cosmic rays above the GKZ cutoff, which is an attractive possibility. However, a more careful analysis is needed in order to verify this, which is to be done in the next subsection.

\subsection{Estimation of the mean life time}
The formula (\ref{ohohoho}) suggest that one may  approximate the decay rate for the diagram \ref{higgs_quark} by the following expression 
\begin{equation}\label{ussr}
\Gamma_{A^0}=y^2\alpha_{h}^4\bigg(\frac{m_p}{m_F}\bigg)^2 m_{A^0}   \ln^2\left(\frac{m_F}{m_p}\right),
\end{equation}
with $m_F$ the mass of a hidden massive fermion in our model and that $m_p$ the mass of the products of the decay. Depending on the case, the coupling constant $\alpha_h$ may refer to the  $\hbox{U(1)}$ 
or the $\hbox{SU(2)}$ interaction.

It is important to remark that formula (\ref{ussr}) is a lower bound for the real decay rate. This follows from the fact that for the Kim diagram the gluons are massless while the hidden $G_\mu^i$ vector bosons in our case are massive, and no correction due to the masses has been introduced in (\ref{ussr}). But it is expected that the mean lifetime will grow for massive $G_\mu^i$ vector bosons, since the decay is less probable when the internal lines corresponds to massive particles. Thus the mean lifetime to be calculated below is smaller than the real one. If the result is larger than the age of the universe, so it will be the real one and the $A^0$ becomes extremely stable. 

The mass of the $A^0$ particle will be chosen between $10^{13}-10^{15}$GeV. This can be achieved by choosing some of the parameters of the potential $Z_i$ or $Y_i$ to have values between $10^{-3}-10^{-5}$. On the other hand a reasonable choice for the coupling constant corresponding to the $\hbox{SU(2)}$ interaction is $\alpha_h\sim 10^{-6}$.This assumed value has the same order of the electroweak interaction, and the hidden interaction is expect  to be of the order or smaller than the weaker known interaction. With this choice of the coupling constant, the mass of the gauge bosons is $f_a g_h\sim 10^{16}$GeV, thus it is larger than the mass of the decaying particle $A^0$, which is the requirement for the ABJ diagram not to be the lowest order one. We can can consider weaker values for $\alpha_h$ without spoiling this condition.  We take also $y\sim 1$, although the Yukawa coupling in our model are usually much smaller. Thus, we are underestimating the real mean life time. In addition, it will be assumed that the super massive particles of the model are of the same order $M\sim 10^{15}-10^{16}$ GeV, with $A^0$ the lowest mass state. This values can be achieved by choosing for instance $v\sim 10^{19}$ GeV,  and the couplings $\lambda_i, Z_i\sim 10^{-3}$ respecting this hierarchy. For this values for the $v_i$ the axion constant is $f_a\sim 10^{19}$ GeV. With these numbers in mind, the formula (\ref{ussr}) gives a mean lifetime
\begin{equation}
\label{lt2}
\tau\sim 10^{14}\; \hbox{yrs}
\end{equation}
which is of three orders larger than the estimated age of the universe. This makes plausible the existence of supermassive particles, whose masses are of the order of $10^{15}\;\hbox{GeV}$ and are present at our era. The products of the decay of such particle may arise as late high-energy cosmic rays above the GKZ limit.
\\

\begin{figure}[!t]
	\begin{center}
		\begin{tabular}{cccccccccccccccc}    %Adds a bunch of centered Columns
			\begin{fmffile}{higgs_quark} 	%one.mf will be created for this feynman diagram
				\fmfframe(6,42)(6,42){ 	%Sets dimension of Diagram
					\begin{fmfgraph*}(220,124) %Sets size of Diagram
						\fmfleft{i}	%Sets there to be 2 sources
						\fmfright{o1,o2}    %Sets there to be 2  outputs
						\fmflabel{$f$}{o1}
						\fmflabel{$f$}{o2}
						\fmf{scalar,label=$A^0$,label.side=left}{i,iv3}
						\fmf{fermion}{ov1,o1}
						\fmf{fermion}{o2,ov2}
						\fmf{fermion,label=$F$,label.side=right}{iv1,iv2,iv3,iv1}
						\fmf{fermion,label=$f$,label.side=left}{ov2,ov1}
						\fmf{boson,label=$G$,lab.dist=0.05w}{iv1,ov1}
						\fmf{boson,label=$G$,label.side=left}{iv2,ov2}
						\fmffixed{(0,.7h)}{iv1,iv2} \fmffixed{(0,.7h)}{ov1,ov2}
					\end{fmfgraph*}	}
			\end{fmffile}		\end{tabular}
		\caption{\textsf{The main decay channel of the hidden Higgs. The gauge fields $G_\mu$ correspond to a very weak interaction with the ordinary matter. This will be the lowest order decay diagram only if the  masses corresponding to the fermion triangle  and the $G_i$ bosons are larger than the $m_{A^0}$. The triangle is composed by a heavy hidden fermion $F$ and the products $f$ of the decay are ordinary fermions of the Standard Model, or hidden fermions with masses close to the light neutrino one $m_\nu$.}}
		\label{higgs_quark}
	\end{center}
\end{figure} 

In any case, although the stability matters of the $A^0$ particle are  non trivial, the discussion made above suggest the existence of scenarios in which this particle is long lived and can be seen in events above the GKZ bound.  

\section{The schizon part of the model}

After ensuring the presence of a super heavy sector with a particle $A^0$ which is stable, the next task is to elaborate the schizon like lagrangian $\mathcal L_{schizon}$ in (\ref{mod2}).  There are several possibilities to be considered. However, the choice should be done with care, since the components of this lagrangian should not spoil the stability just achieved. To give an example of this situation, note that in the previous section we considered a that the double Higgs mechanism gives mass to the gauge bosons, and therefore the Goldstone boson $G^0$ is eaten by a gauge transformation and is not physical. This was done intentionally, otherwise, the axion $G^0$ will be coupled directly or indirectly to the $A^0$ particle with a coupling of the form $A^0 (G^0)^3$. This induces a decay of the form 
$$
A^0\to G^0+G^0+G^0.
$$ 
This decay will make $A^0$ short lived. For this reason choose an scenario which give mass to the vector bosons $G_\mu^i$, which avoids such coupling and simultaneously insuring the stability of $A^0$.

In order for an extremely light axion $a$ to emerge a $U(1)$ symmetry breaking mechanism at the Planck scale is required. Several possibilities for doing this will be considered below.
One of them is to consider an additional fermion $Q$ which acquires mass 
through an spontaneously symmetry breaking with an scalar $\varphi$, which acquires an expectation value $\varphi_0\sim f_a\sim M_{pl}$. This fermion $Q$ is therefore super massive. The fermion $Q$ should participate in a gauge interaction in such a way that the $U(1)$ current $J_\mu^5$ is anomalous. If this interaction involves light fermions $x$, with masses $m_x$ of the order of the MeV scale, which have non zero condensates $<\overline{x} x>$, then the axion may obtain the required mass.  The reasoning is analogous to the one giving (\ref{vives}).

The next task is to identify the interaction giving rise to such condensates $<\overline{x} x>$. The simplest possibility is to consider the gauge interaction as the $\hbox{SU(2)}_L$ described in previous section. The axion will acquire the desired mass if there are no condensates of the form $<\overline{F}_iF_i >$, otherwise the axion mass will be 
$$
m^2_a\sim\frac{M_{GUT}}{f^2_a} <\overline{F}_iF_i >,
$$ 
which is unacceptably large for our purposes. Thus, this may happen if $<\overline{F}_iF_i >=0$ and  if there exist condensates $<\overline{x}x>$  corresponding to particles of the Standard Model which extremely tiny values, of the order of the neutrino mass $m_\nu$. If the axion were the QCD one and the scale of symmetry breaking is the Planck scale then its mass would be
$$
m_a\sim \frac{f_\pi m_\pi}{f_a}\sim 10^{-9} eV,
$$
which is 23 orders of magnitude larger than required. However, this calculation is done in the QCD context. If instead, the interaction is the hidden one, this formula should be corrected by a factor
$$
m_a\sim \frac{f_\pi m_\pi}{f_a}\sqrt{\frac{<\overline{X}X>_{h}}{<\overline{X}X>_{\mathrm{QCD}}}},
$$
due to the change from the QCD to the hidden interaction. The orders of magnitude are corrected if \cite{Calogero}-\cite{Estrada}
$$
\frac{<\overline{X}X>_{h}}{<\overline{X}X>_{\mathrm{QCD}}}\sim 10^{-23}\sim \frac{m_\pi}{M_{Pl}}.
$$
This is the case when the theory is confining at an scale $\Lambda_h\sim 10^{-4}$eV. A more attractive possibility is the existence of particles $x$ with masses $m_x\sim m_\nu$ and  that the confining scale $\Lambda_h\sim m_\nu$ as well. These scales work for our purposes by virtue of (\ref{vives}). These particles may be ordinary neutrinos $\nu$ or hidden particles. It is important to remark that the addition of such light particles do not spoil the stability of $A^0$ since, as discussed in previous section, the lighter of the products of the decay are, the larger of the mean life time of  $A^0$ becomes. A lagrangian with these characteristics is
$$
\mathcal L_{schizon}=\mathcal L_{Q ssb}(Q, \varphi)+i\sum_{i=1}^N \overline{f}_{iL} \slashed{\nabla} f_{iL} +i\sum_{i=1}^N \overline{f}_{iR} \slashed{\partial} f_{iR} 
-(\delta^i \overline{f}_{iR} \Phi_3f_{iL} +\delta_i^\ast \overline{f}_{iL} \Phi_3^\ast f_{iR}) 
$$
$$
-\kappa^i \Phi_4 f_R^T Cf_R+V(\Phi_3)+V(\Phi_4).
$$
The lagrangian $\mathcal L_{Q ssb}$ contains a heavy fermion $Q$ which plays the role of the hidden quark in the KSVZ scenario. The $Q$ fermion acquires
a mass with a neutral complex Higgs $\varphi$. This part of the lagrangian is assumed to be invariant under a Peccei-Quinn like symmetry
$$
Q_L\to e^{i\alpha} Q_L,\qquad Q_R\to e^{i\alpha} Q_R,\qquad a\to a+2f_a \alpha
$$
where the axion $a$ as usual is related to the phase of $\varphi$. This symmetry is anomalous and thus the axion $a$ acquires a mass due to 
the small masses of the new fermions $f_i$. In fact, a simple inspection shows that the $f_i$ fermions have Dirac and Majorana mass terms.  The potentials $V(\Phi_i)$ are have the standard form
$$
V(\Phi_i)=m^2\Phi_i^\ast\Phi_i -\lambda(\Phi_i^\ast\Phi_i)^2,
$$
thus the field $\Phi_i$ acquire a non zero expectation value. The mass term for $f_i$ is then
 \begin{equation*}
	\overline{\overline{\mathbf{M}}}_i=
	\begin{pmatrix*}
	0 & m_{if}  \\
	m_{if}  & M_{if}
	\end{pmatrix*}  \;,
\end{equation*}
where $m_{if}=\delta^i v_1$ and $M_f= \kappa^i v_3$, with $v_i$ the minimum of the potential for $\Phi_i$. If $M_f\gg m_f$, which means that the Majorana terms 
give the greatest contribution, then the approximate eigenvalues are (we omit the index $i$ in order to make the notation more readable)
\begin{equation*}
\lambda^f_1\sim \frac{m_f^2}{M_f}	\qquad \hbox{and} \qquad \lambda^f_2 \sim M_f   \;,
\end{equation*}
and the mass eigenstates are
\begin{equation*}
\begin{gathered}
	f_1\simeq \frac{1}{\sqrt{1+\bigg(\frac{M_f}{m_f}\bigg)^2}}f_R+\frac{\frac{M_f}{m_f}}{\sqrt{1+\bigg(\frac{M_f}{m_f}\bigg)^2}}f_L  \;, \\
	f_2\simeq \frac{1}{\sqrt{1+\bigg(\frac{M_f}{m_f}\bigg)^2}}f_L-\frac{\frac{M_f}{m_f}}{\sqrt{1+\bigg(\frac{M_f}{m_f}\bigg)^2}}f_R  \;.
\end{gathered}
\end{equation*}
Correspondingly, if the previous mass limit is satisfied then $f_1\sim f_L$ and $f_2\sim f_R$.  This is an example of a seesaw mechanism. For instance, by choosing  $\Phi_i$ in order to get $m_f\sim $TeV and $M_u\sim M_f\sim 10^{15}$ GeV it follows that the small mass eigenvalue is $\lambda_1\sim m_\nu$ and $\lambda_2 \sim 10^{15}$ GeV. Thus, this choice cast a very light sector with masses of the order of the mass of a neutrino $m_\nu$, which is one of our requirements. The heavy masses are required to be slightly larger than the mass of the $A^0$ particle in order to do not introduce new decay channels for this particle.

We arrive then to the conclusion that the model constructed above possess fermions with mass $m_\nu$, as desired. If these particles posses a condensate $<\overline{f}_L f_L>\sim m_\nu^3$ the axion $a$ coming from the $Q$ sector will obtain the required mass to model a considerable fraction of the energy density of the universe (\ref{vives}).

A comment about the axion $a$ described above is in order. The seesaw mechanism implies that the light mass eigenstates are predominantly the left ones, but with a small mixture with the right states. Thus the massive states are also interacting through the $\hbox{SU(2)}_L$ gauge interaction. This may introduce a potentially dangerous correction to the axion mass (\ref{vives}) of the form
$$
m^2_a\sim\frac{\lambda_2 \Lambda^3_h}{f^2_a} \;,
$$
with $\lambda_2\sim 10^{13}$ GeV  the large mass eigenvalue discussed above. This makes the axion mass very large for our purposes. However, after careful analysis of the mixing, it is obtained that the left part of the  lagrangian written in terms of the mass eigenstates is schematically
$$
L_{left}= \overline{f}_{L} \slashed{\nabla} f_{L} \sim \overline{f}_{1} \slashed{\nabla} f_{1} -\frac{\lambda_1}{\lambda_2}(\overline{f}_{1} \slashed{\nabla} f_{2} +\overline{f}_{2} \slashed{\nabla} f_{1})+\bigg(\frac{\lambda_1}{\lambda_2}\bigg)^2 \overline{f}_{2} \slashed{\nabla} f_{2},
$$
where $f_1$ is the light mass eigenstate; $f_2$ the massive one and $\lambda_i$ the corresponding mass eigenvalues. Thus, it can be seen that the last term induces an axion mass of the form
$$
m^2_a\sim\bigg(\frac{\lambda_1}{\lambda_2}\bigg)^2\frac{\lambda_2 \Lambda^3_h}{f^2_a}\sim \frac{\lambda^2_1 \Lambda^3_h}{\lambda_2 f^2_a}  \;.
$$
Since $\lambda_1\sim m_\nu$ and $\lambda_2\sim 10^{15}$GeV it follows that the last expression several orders smaller than (\ref{vives}). The mixed terms $\overline{f}_{1} \slashed{\nabla} f_{2} $ induce a mass of the form
$$
m^2_a\sim\frac{\lambda_1}{\lambda_2} \frac{\sqrt{\lambda_2\lambda_1} \Lambda^3_h}{f^2_a} \sim\frac{\lambda^{3/2}_1 \Lambda^3_h}{\sqrt{\lambda_2}f^2_a}  \;.
$$
which is again of smaller order as (\ref{vives}). Thus, the large contributions to the axion mass seem to be cancelled due to the smallness of the mixing terms.  Thus, if heavy fermion condensates are absent, this model may cast such extremely light axions. The renormalization group for the coupling constant $g_h$ for the $\hbox{SU(2)}_L$ theory should be such that the mass $m_{G}\sim g_h f_a$ of the bosons $G_\mu^i$ do not become smaller that the mass of the $A^0$ at the scales required at figure  \ref{higgs_quark}.

Of course, there is no need to consider the $\hbox{SU(2)}_L$ gauge interaction as the only possible source of the axion. One may consider that the axion $a$ acquires an small mass due to an anomaly corresponding to an interaction very weak, of the gravitational order. However, the choice of  $\hbox{SU(2)}_L$ as the responsible for the axion is minimal, and these possibilities are always more attractive.

A further possibility is to consider schizons of the neutrino type as in \cite{frieman}-\cite{frieman3}. In these models there are $N$ light fermions with mass terms of the form
\begin{equation}\label{neutroso}
\mathcal L_{schizon}=\frac{1}{2}\partial_\mu a\partial^\mu a+i\sum_{j=1}^N \overline{f}_{jL} \slashed{\nabla} f_{jL} +i\sum_{j=1}^N \overline{f}_{jR} \slashed{\partial} f_{jR} 
+\bigg(m_0+\epsilon\exp{i (\frac{a}{f_a}+\frac{2\pi  j}{N})}\bigg)\overline{f}_{jL} f_{jR}+h.c
\end{equation}
The small masses may come from a see saw mechanism of the type described above, and this explains the presence of the large scale $f_a\sim M_{pl}$ in the model.
Strictly speaking the eigenstates of masses are not $f_L$ or $f_R$, but an small mixture. However, since the see saw mechanism involves an extremely heavy scale, we will neglect the effect of this mixture.
In the same fashion as above, we consider masses $m_0\sim \epsilon\sim m_\nu$.  
The lagrangian (\ref{neutroso}) has the approximate symmetry
$$
f_j\to f_{j+1},\qquad f_N\to f_1,\qquad a\to a+2\pi j f_a/N,
$$
which is broken by the $\epsilon$ term. The 1-loop induced potential for the pseudoscalar $a$ is
 $$
 V(a)=\sum_{i=1}^N \frac{M^4_i}{16\pi^2} \log \frac{\Lambda^2}{M^2_i}
 $$
 where $M_i^2$ is given by
 $$
 M_i^2=m_0^2+\epsilon^2+2m_0 \epsilon \cos\bigg(\frac{a}{f_a}+\frac{2\pi j}{N}\bigg)
 $$
Thus  the axion mass becomes of the order
$$
m_a\sim \frac{m_0 \epsilon}{f_a}\sim 10^{-32}eV,
$$
which is of the desired order. Note that the light masses do not spoil the stability of the $A^0$ particle, neither the heavy ones.

In brief, we have strongly suggested some scenarios which can cast super heavy particles which are stable, even in presence of extremely light states. The double Higgs mechanism is 
essential in this construction, since it posses a pseudo scalar $A^0$ whose stability is more easy to handle that in the scalar case.

\section{Discussions and open perspectives}

In the present work the possibility of modeling the vacuum energy density of the present universe in terms of the potential energy of an axion pseudo scalar $a$ was considered. The hypothetical axion of the model presented here is not identified with a QCD invisible one; instead, it is a pseudo-Goldstone boson corresponding to a global symmetry which is spontaneously broken at a scale $f_a$ of the order of the Planck mass $M_{Pl}$. The hidden axion acquires a potential due to the presence of  small mass terms which violates explicitly this symmetry. The sector possess a symmetry  $\hbox{SU(2)}_\mathrm{L}$ gauge interaction which is confining at an energy scale close to a light neutrino mass $m_{\nu}\sim 10^{-2}$eV. The axion mass arising from the generated potential is  $m_a\sim 10^{-32}\;\hbox{eV}$, thus it becomes the lightest massive particle in the universe. Its Compton wavelength is of the order of the Hubble radius and therefore this component represents a considerable fraction of the critical energy density $\rho_c$, since its relaxation time is of the order of the age of the present universe. 

The hidden scenario yielding this axion is composed by supermassive and very light particles, whose masses arise from a spontaneous symmetry breaking together with a suitable seesaw mechanism, which induces very large and very tiny mass eigenvalues for the fermion sector. The tiny masses violate a global symmetry present in the super heavy sector and are needed in order to generate the axion potential. In addition, this scenario suggest the existence of a stable superheavy particle, a hidden pseudo scalar $A^0$, with a mass $m_{A^0}\sim 10^{15}\;\hbox{GeV}$, which may be present at our era. The particles composing this hidden sector have mass values which matches the ones described in \cite{riotto}, that is, weakly interacting matter with masses near or above the order of the GUT scale, see also \cite{riotto2}. A proof on whether such supermassive particles exist or not in the present universe is beyond the scope of this work. There are discussions in the context of supersymmetric models that suggest that they may exist \cite{suheavy}. However, our work is focused in characterizing the mechanisms which, if these hypothetical particles do exist, guarantee their stability. An interesting feature is that such massive particles may generate events above the GKZ bound at the present universe and this is a possibility to be analyzed further.

To find a common scenario possessing all these features at once, as showed in the text, is a non trivial task. The model presented here assumes that the quantum field theory vacuum energy contributions are not present, perhaps due to a suitable adjustment mechanism such as \cite{dolgov,dolgov2,urban,emelyanov,emelyanov1,emelyanov2}. This is standard in quintessence like mechanisms.

 It may be a valuable task to identify supersymmetric models or superstring compactifications which encompass scenarios of the type presented here. In fact, double Higgs mechanisms appear in some supersymmetric extensions of the Standard Model \cite{gunion}, and between these models interesting candidates may emerge. It may also be of interest to cast scenarios of the type presented here in the context of technicolor models \cite{technicolor}-\cite{technicolor}. There is an interesting discussion about cosmological applications of such models in \cite{olinto}, but the discussion of that work is related to inflationary problems, not to the current accelerated expansion of the universe. So, to construct technicolor inspired models with extremely tiny mass axions and super heavy stable particles may be an interesting task to achieve. We leave these matters for a future investigation.

\vspace{20pt}{\bf Acknowledgements:}  An important discussion about the double Higgs mechanism with Prof. Ezequiel Alvarez is warmly acknowledged. This work is dedicated to Luis Masperi, who was working in similar subjects before his decease. O.S is supported by CONICET (Argentina).


\begin{thebibliography}{99}

\bibitem{Rubin} V. Rubin and W. Ford, \textit{Astrophysical Journal} \textbf{159} (1970) 379.
\bibitem{Rubin2} V. Rubin, D. Burstein, W. Ford Jr and N. Thonnard, \textit{Astrophysics J.} \textbf{289} (1985) 81.

\bibitem{Carroll} S. Carroll, W. H. Press and E. Turner, \textit{Annu. Rev. Astron. Astrophys.} \textbf{30} (1992) 499.

\bibitem{Salam} A. Salam, \textit{Nuovo Cimento} \textbf{5} (1957) 299.    
\bibitem{Lee} T. Lee and C. Yang, \textit{Phys. Rev.} \textbf{104} (1956) 254.
\bibitem{mirror} L. Okun, \textit{Phys. Usp.} \textbf{50} (2007) 380.
\bibitem{hidden33} J. Feng, H. Tu and Hai-Bo Yu, \textit{JCAP} \textbf{0810} (2008) 043.
\bibitem{hidden2} J. Espinosa, T. Konstandin, J.M. No and M. Quiros, \textit{Phys. Rev. D} \textbf{78} (2008) 123528.
\bibitem{wicax} B. Patt and F. Wilczek, ``Higgs-field Portal Into Hidden Sectors", hep-ph/0605188.
\bibitem{Higgs1} O. Bertolami and R. Rosenfeld, \textit{Int. J. Mod. Phys. A} \textbf{23} (2008) 4817.                  %CITA 10
\bibitem{Higgs2} J. March-Russell, S. West, D. Cumberbatch and D. Hooper, \textit{JHEP} \textbf{0807} (2008) 058.
\bibitem{Higgs3} C. Englert, T. Plehn D. Zerwas and P. Zerwas, \textit{Phys. Lett. B} \textbf{703} (2011) 298.
\bibitem{Higgs5} O. Lebedev and H. Lee, \textit{Eur. Phys. J. C} \textbf{71} (2011) 1821.
\bibitem{Higgs6} R. Dick, R. Mann and K. Wunderle, \textit{Nucl. Phys. B} \textbf{805} (2008) 207.   

\bibitem{quintesence} S. Carroll, \textit{Phys. Rev. Lett.} \textbf{81} (1998) 3067.

\bibitem{dolgov} A. Dolgov, \textit{Phys. Rev. D} \textbf{55} (1997) 5881; V. Rubakov and A. Tinyakov, \textit{Phys. Rev. D} \textbf{61} (2000) 087503.
\bibitem{dolgov2}  A. Dolgov and M. Kawasaki, \textit{Physics of Atomic Nuclei} \textbf{68}, Issue 5 (2005) 828.
\bibitem{urban} A. Dolgov and F. Urban, \textit{Phys. Rev. D} \textbf{77} (2008) 083503.
\bibitem{emelyanov} V. Emelyanov and F. Klinkhamer, \textit{Int. J. Mod. Phys. D} \textbf{21} (2012) 1250025.
\bibitem{emelyanov1} V. Emelyanov and F. Klinkhamer, \textit{Phys. Rev. D} \textbf{85} (2012) 103508.                      %CITA 20
\bibitem{emelyanov2} V. Emelyanov and F. Klinkhamer, \textit{Phys. Rev. D} \textbf{86} (2012) 027302.

\bibitem{dolgovv} A.D. Dolgov, ``The Very Early Universe", Proc. of the Nuffield Workshop, (Eds. G. Gibbons, S.W. Hawking and S.T. Tiklos), Cambridge University Press, 1982, 449.

\bibitem{weinberg} S. Weinberg, \textit{Rev. Mod. Phys.} \textbf{61} (1989) 1.

\bibitem{GKZ} K. Greisen, \textit{Phys. Rev. Lett.} \textbf{16} (1966) 748; G. T. Zatsepin and V. A. Kuzmin,
\textit{Sov. Phys.-JETP Lett.} \textbf{4} (1966) 78.

\bibitem{belavin} A. Belavin, A. Polyakov, A. Schwarz and Y. Typkin, \textit{Phys. Lett. B} \textbf{59} (1975) 85.

\bibitem{thooft} G. 't Hooft, \textit{Phys. Rev. Lett.} \textbf{37} (1976) 8. G. 't Hooft, \textit{Phys. Rev. D} \textbf{14} (1976) 3432.
 

\bibitem{axion1} R. Peccei and H. Quinn, \textit{Phys. Rev. Lett.} \textbf{38} (1977) 1440; R. Peccei and H. Quinn, \textit{Phys. Rev. D} \textbf{16} (1977) 1791.
 

\bibitem{Kim1987} J. Kim, \textit{Phys. Rept.} \textbf{150} (1987) 1.                             %Cita 30

\bibitem{axion11} F. Wilczek, \textit{Phys. Rev. Lett.} \textbf{40} (1978) 279.


\bibitem{axion12} S. Weinberg, \textit{Phys. Rev. Lett.} \textbf{40} (1978) 223.
\bibitem{axion3} W. Bardeen and S. Tye, \textit{Phys. Lett. B} \textbf{74} (1978) 229.
\bibitem{axion4} D. Chang and R. Mohapatra, \textit{Phys. Rev. D} \textbf{32} (1984) 293.
\bibitem{axion5} M. Dine, W. Fischler and M. Srednicki, \textit{Phys. Lett. B} \textbf{104} (1981) 199.
\bibitem{axion6} D. Kaplan, \textit{Nucl. Phys. B} (1985) 260.
\bibitem{axion7} W. Bardeen, R. Peccei and T. Yanagida, \textit{Nucl. Phys. B} (1987) 401.
\bibitem{axion7bis} M. Srednicki, \textit{Nucl. Phys B} \textbf{260} (1985) 689.


\bibitem{SVZ1980} M. Shifman, A. Vainstein and V. Zakharov, \textit{Nucl. Phys. B} \textbf{166} (1980) 493.

\bibitem{kim1979} J. Kim, \textit{Phys. Rev. Lett.} \textbf{43} (1979) 103.

\bibitem{Calogero} F. Calogero, \textit{Phys. Lett. A} \textbf{238} (1997) 335.

\bibitem{Estrada} J. Estrada Vigil and L. Masperi, \textit{Mod. Phys. Lett. A} \textbf{13} (1998) 423.

\bibitem{frieman} J. Frieman, C. Hill and R. Watkins, \textit{Phys. Rev. D} \textbf{46} (1992) 1226.

\bibitem{frieman3} C. Hill and G. Ross Nucl. Phys. B 311 (1988) 253. C. Hill and G. Ross Phys. Lett. B 203 (1988) 125.

\bibitem{caldwell} R.R. Caldwell and M. Kaminkoeski, \textit{Ann. Rev. Nucl. Part. Sci.} \textbf{59} (2009) 397.


\bibitem{quv} J. Kim and H.P. Nilles, \textit{Phys. Lett. B} \textbf{553} (2003) 1.


\bibitem{riotto} E. Kolb, D. Chung and A. Riotto, ``DARK98", Proceedings of the Second International Conference on Dark Matter in Astro and Particle Physics, Edited by H.V. Klapdor-Kleingrothaus and L. Baudis.
\bibitem{riotto2} E. Kolb, A. Starobinsky and I. Tkachev, \textit{JCAP} \textbf{0707} (2007) 005.              %Cita 40


\bibitem{suheavy} V. Berezinsky, M. Kachelriess and M. Solberg, \textit{Phys. Rev. D} \textbf{78} (2008) 123535.

\bibitem{georgi} H. Georgi, Hadronic J. 1 (1978) 155.

\bibitem{hunter} J. Gunion, H.  Haber, G. Gordon and S. Dawson "The Higgs Hunter's Guide" Addison-Wesley, 1990.

\bibitem{hunter2} H. Haber and D. O' Neil Phys. Rev. D 74 (2006) 015018.

\bibitem{hunter3} S. Davidson and H. Haber Phys. Rev. D 72 (2005) 035004; Erratum-ibid.D 72 (2005) 099902; J. Gunion and H. Haber Phys. Rev. D 72 (2005) 095002; H. Haber and D. O' Neil Phys .Rev. D 83 (2011) 055017.

\bibitem{gunion} J. Gunion and H. Haber Nucl. Phys. B 272 (1986) 1, Nucl. Phys. B 402 (1993) 567; Nucl. Phys. B 278 (1986) 449. 

\bibitem{technicolor}  S. Weinberg Phys. Rev. D 13  (1976) 974; Phys. Rev. D19 (1979) 1277. 

\bibitem{technicolor2} L. Susskind  Physical Review D20 (1979) 2619.

\bibitem{olinto} F. Adams, J.  Bond, K. Freese, J. Frieman and A. Olinto Phys. Rev. D 47 (1993) 426.

\bibitem{nilles} J. Kim and H. Nilles "Dark energy from approximate $U(1)_{de}$ symmetry" arXiv:1311.0012.
\bibitem{nilles2} J.Kim  JHEP 9905 (1999) 022; JHEP 0006 (2000) 016.
\end{thebibliography}
\end{document}